\documentclass[11pt]{article}

\usepackage[preprint]{acl}

\usepackage{times}
\usepackage{latexsym}

\usepackage[T1]{fontenc}

\usepackage[utf8]{inputenc}

\usepackage{microtype}

\usepackage{inconsolata}

\usepackage{graphicx}
\usepackage{booktabs}
\usepackage{multirow}
\usepackage{amsmath}
\usepackage{verbatim}
\usepackage{booktabs}
\usepackage{array}
\usepackage{tabularx}
\usepackage{listings}
\usepackage[inline]{enumitem}
\usepackage{url}
\usepackage{subcaption}
\usepackage{soul}
\usepackage{amsmath}
\usepackage{amsfonts}
\usepackage{tcolorbox}
\usepackage{booktabs}
\usepackage{multicol}
\usepackage{multirow}
\usepackage{algorithm}
\usepackage{algpseudocode}
\usepackage{colortbl}
\usepackage{amssymb}
\usepackage{wrapfig}
\usepackage{svg}
\usepackage{longtable}
\usepackage{xcolor}

\title{FLARE: Fine-Grained Diagnostic Feedback for LLM Code Refinement}

\author{
Yinsheng Yao\thanks{Equal contribution.} \\
Johns Hopkins University \\
Baltimore, MD, USA \\
\And
Hongxiang Zhang\footnotemark[1] \\
Purdue University \\
West Lafayette, IN, USA \\
\texttt{hxxzhang@purdue.edu}
\AND
Weixi Tong \\
Purdue University \\
West Lafayette, IN, USA \\
\texttt{tong172@purdue.edu}
\And
Tianyi Zhang \\
Purdue University \\
West Lafayette, IN, USA \\
\texttt{tianyi@purdue.edu}
}

\usepackage{soul}
\definecolor{lightred}{rgb}{1.0, 0.8, 0.8}
\definecolor{lightblue}{RGB}{255, 255, 255}
\definecolor{lightgreen}{RGB}{210, 240, 210}
\newcommand{\todo}[1]{}

\definecolor{lightgray}{RGB}{225, 225, 225}
\definecolor{lightred}{rgb}{1.0, 0.8, 0.8}

\newcommand{\headercolorlong}{\rowcolor{gray!15}}

\newcommand{\gain}[1]{\textcolor{green!50!black}{\scriptsize$_{\uparrow #1}$}}

\newcommand{\method}{\textsc{Flare}}

\begin{document}
\maketitle

\begin{abstract}
Large language models often generate code with bugs. 
Existing methods revise such code using feedback signals such as test failures and self-critiques, but these signals are too coarse-grained or too high-level to tell the model where the code should be revised. We present {\method}, an iterative code generation framework whose lightweight diagnostic model predicts line-level suspiciousness to guide code refinement.
Given the inherent uncertainty of diagnostic predictions, {\method} searches over the top-k suspicious lines at each step and selects the next candidate according to execution outcomes. 
Experiments on LiveCodeBench and BigCodeBench with five base LLMs show that, even without candidate search, {\method} ($k$=1) outperforms the strongest baseline, NL-Debugging, with an absolute improvement from 1.06\% to 9.71\% in 9 of 10 settings. 
Expanding the search to 10 candidates adds an average improvement of 8.97\%, and {\method} keeps an average lead of 3.39\% over the strongest baseline under the same candidate budget.
When evaluated in isolation, our lightweight diagnostic model outperforms recent fault localization methods, demonstrating that it can provide reliable fine-grained guidance for code refinement.
\end{abstract}
\section{Introduction}
Large Language Models (LLMs) have demonstrated remarkable capabilities in coding~\citep{10.1145/3747588,achiam2023gpt,zhu2024deepseek}. 
Despite this progress, LLMs remain prone to generating code with bugs \citep{jesse2023large, wang2025towards}.
Given the challenge of generating correct code in one pass, existing code generation methods often perform iterative refinement, where a model generates an initial program, receives feedback, and then refines the program over multiple rounds~\citep{li-etal-2025-test,chen2023teaching,madaan2023selfrefine,zhang-etal-2023-self,zhang2025nldebugging}. 
However, these methods usually rely on feedback signals such as test failures and self-critiques to iteratively refine the generated code. Such signals are either too coarse-grained or too high-level to tell the model where the code should be revised.

To address this limitation, we propose {\method}, an iterative framework that uses a lightweight diagnostic model to predict line-level suspiciousness for targeted refinement during code generation. 
Given an initial code solution, {\method} aggregates the language model's subword-token probabilities over AST-aligned lexical units such as identifiers, operators, and keywords.
The diagnostic model scores these lexical units and aggregates the scores into line-level suspiciousness predictions. 
During refinement, {\method} combines the suspicious line with execution feedback to generate targeted candidate refinements. 
Given the inherent uncertainty of fine-grained feedback predictions, {\method} is further equipped with a search procedure over the top-$k$ suspicious lines at each iteration and selects the next candidate based on execution outcomes. 
Note that we do not assume that the program contains only one bug. Instead, {\method} addresses one bug at a time and iteratively refines
the program.

\begin{figure*}[t]
    \centering
    \includegraphics[width=\textwidth]{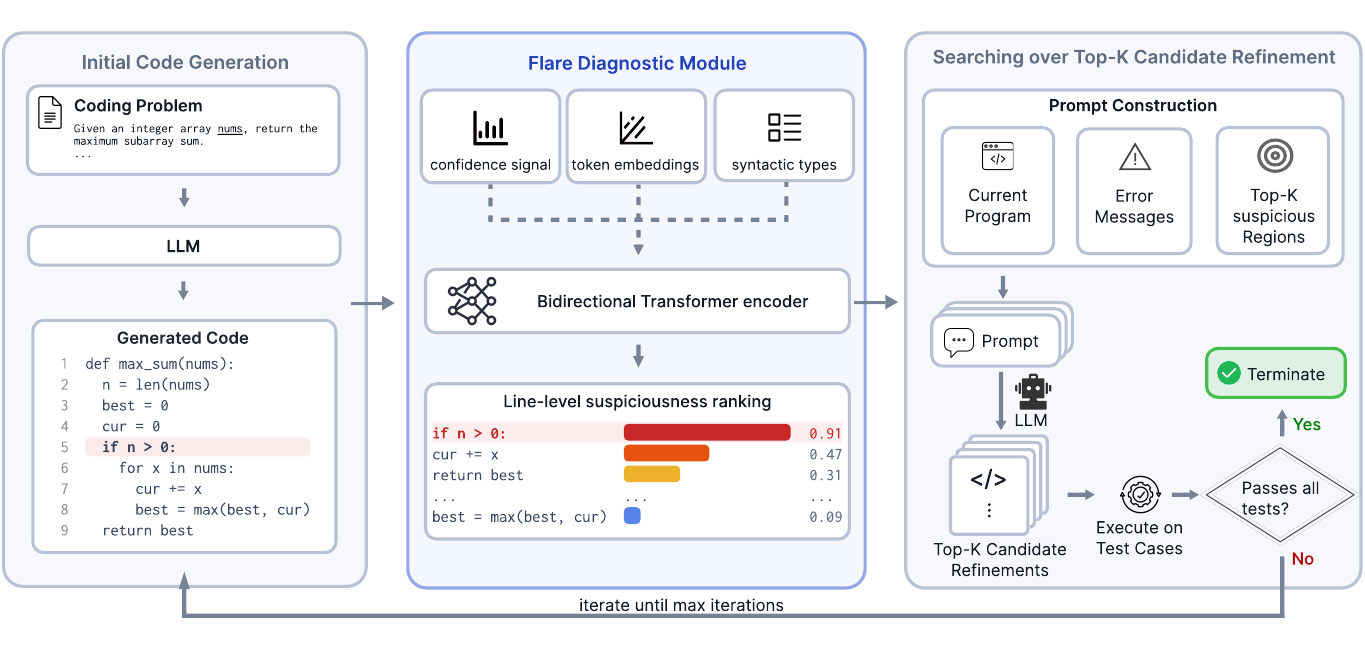}
    \caption{Overview of {\method}. The framework aligns LLMs' probability signals to predict line-level suspiciousness, combines this diagnostic signal with execution feedback, and searches over top-$k$ candidate refinements.}
    \label{fig:framework}
\end{figure*}

We evaluate {\method} on two code generation benchmarks,  LiveCodeBench~\citep{jain2024livecodebench} and BigCodeBench~\citep{zhuo2024bigcodebench},  with five base LLMs and three existing iterative refinement methods, including NL-Debugging~\citep{zhang2025nldebugging}, Self-Debugging~\citep{chen2023teaching}, and Self-Refine~\citep{madaan2023selfrefine}. 
Even without candidate search ($k=1$), {\method} outperforms the most competitive baseline, NL-Debugging, in 9 of 10 settings with an absolute improvement from 1.06\% to 9.71\%. 
Furthermore, searching over 10 candidates ($k=10$) yields an absolute average improvement of 8.97\% compared with no candidate search. Under the same $k=10$ budget, {\method} still outperforms NL-Debugging and Best-of-N~\citep{snell2024scaling} by 3.39\% and 10.45\% absolute improvement on average, respectively. This suggests that diagnostic localization complements candidate search by using the expanded budget more effectively.
Ablation studies further show that both fine-grained diagnostic feedback and candidate search contribute significantly to overall gains. 
When evaluated in isolation, our diagnostic model achieves 67\% Top-1 and 89\% Top-10 localization accuracy, outperforming recent fault localization methods, including FlexFL~\citep{flexfl}, LLMAO~\citep{yang2024large}, and BAP~\citep{stein2025where}. 

\section{Methodology}
    \label{sec:methodology}
    Figure~\ref{fig:framework} provides an overview of {\method}. 
    Given a code generation task and an initial program generated by an LLM, {\method} predicts ranked suspicious lines with a lightweight diagnostic model  (Section~\ref{sec:fault_localization}). 
    {\method} then prompts the LLM to generate revisions for the top-$k$ suspicious lines and selects the next program according to execution outcomes (Section~\ref{sec:bon}).
    The process continues until a program passes all tests or reaches a maximum iteration budget.

\subsection{Fine-Grained Diagnostic Feedback}
    \label{sec:fault_localization}

    Existing self-refinement methods for code generation often rely on coarse-grained or high-level feedback, which provides limited guidance about where the underlying issue resides. 
    {\method} introduces a fine-grained diagnostic model that ranks suspicious lines of the generated program before each refinement step.\footnote{We evaluate the diagnostic model and ablate its components in Section~\ref{sec:diagnostic}.} 
    Formally, given a generated program $\mathcal{C}$ segmented into $L$ lines $\mathcal{C}=[c_1,\dots,c_L]$, 
    we train a diagnostic model $f_\theta$ that outputs line-level suspiciousness scores: 
    \begin{equation}
        \mathbf{r}=f_\theta(\mathcal{C}) \in \mathbb{R}^{L},
        \qquad
        \mathbf{r}=[r_1,\dots,r_L],
    \end{equation}
    where $r_i$ denotes the suspiciousness score assigned to line $c_i$. 
    

    \paragraph{Training Data Construction.}
        \label{sec:data_collection_input_representation}
        We train the diagnostic model on the training split of \texttt{Collu-Bench}~\citep{jiang2024collubench}, which contains 10504 buggy code solutions generated by 11 diverse LLMs on five datasets, including HumanEval, MBPP, HumanEval-Java, Defects4J, and SWE-bench.\footnote{None of the benchmarks used in our evaluation is included in \texttt{Collu-Bench}.} Each solution is paired with token probabilities from the generator LLM and an annotation of the first buggy token. 
        Previous empirical studies reveal two key insights~\citep{jiang2024collubench, spiess2025calibration}: LLMs tend to exhibit lower confidence when generating buggy code, and their errors are concentrated on certain token categories, such as \texttt{Keyword}, \texttt{Identifier}, and \texttt{Type}. Motivated by these findings, we represent each lexical unit by its embeddings, syntactic type, and the LLM's confidence signals derived from subword-token probabilities.

        To construct these inputs, we first align the LLM's confidence signals with the code's lexical-unit representation. While the LLM assigns probabilities to subword tokens, program semantics are organized around syntactic units such as identifiers, operators, and statements. This mismatch is particularly problematic for code, as standard Byte-Pair Encoding (BPE) tokenization often splits a single lexical unit into multiple subword pieces, e.g., \texttt{max\_retries} $\rightarrow$ $\{$\texttt{max}, \texttt{\_}, \texttt{retries}$\}$.

        Specifically, we parse $\mathcal{C}$ and extract the AST-aligned lexical sequence
        \begin{equation}
            \mathcal{T}(\mathcal{C})=\{t_1,\dots,t_M\},
        \end{equation}
        where $t_j$ corresponds to a lexical unit.

        Let $\{p_n\}_{n=1}^{N}$ denote the probabilities of the $N$ subword tokens. For a lexical unit $t_j$, let $\mathcal{S}(j)\subseteq\{1,\dots,N\}$ contain the indices of its aligned subword tokens. We define its aggregate confidence
        \begin{equation}
            p'_j = \prod_{n\in \mathcal{S}(j)} p_n.
            \label{eq:lexical_confidence}
        \end{equation}

        $p'_j$ represents the joint probability assigned by the language model to lexical unit $t_j$ and serves as a measure of the model's confidence in that unit. 
        Because this product may be sensitive to the number of aligned tokens, Appendix~\ref{app:length_adjustment} compares it with alternative aggregation methods that eliminate this effect.

        For each lexical unit $t_j$ in the program $\mathcal{C}$, we represent it by the tuple
        \begin{equation}
            \mathbf{x}_j =
            \bigl(
            t_j,\;
            \mathrm{Type}(t_j),\;
            p'_j
            \bigr),
            \label{eq:input_rep}
        \end{equation}
        where $\mathrm{Type}(t_j)$ denotes the syntactic category of $t_j$, such as \texttt{Keyword}, \texttt{Identifier}, or \texttt{Type}.\footnote{We list type details in Appendix~\ref{app:type-inventory}.}     

    \paragraph{Training Objective.}
        \label{sec: Fault Localization}
        Diagnosing a bug often requires context beyond one lexical unit or line.
        For example, a variable update or branch condition may only appear suspicious when compared with definitions, uses, or constraints elsewhere in the program. 
        To model these dependencies, we use a bidirectional transformer encoder to encode the AST-aligned sequence, allowing each lexical unit to attend to the full program context~\citep{guo2021graphcodebert}.


        Given the program $\mathcal{C}$, we first embed each lexical unit $t_j$ in $\mathcal{C}$ into an input vector by concatenating its token embedding, syntactic type, and projected confidence:
        \begin{equation}
            \begin{aligned}
            \mathbf{e}_j
            &=
            \bigl[
            \mathrm{Emb}_{tok}(t_j);\,
            \mathrm{Emb}_{type}(\mathrm{Type}(t_j));\\
            &\qquad
            \mathrm{Proj}(p'_j)
            \bigr],
            \end{aligned}
        \end{equation}
        where $\mathrm{Emb}_{tok}$ and $\mathrm{Emb}_{type}$ are learned embedding layers, and $\mathrm{Proj}$ maps the confidence feature $p'_j$ to a vector representation.\footnote{The implementation details are provided in Appendix~\ref{appendix: exp details}}

        We add positional encodings and feed the resulting sequence into the transformer encoder to obtain contextual representations:
        \begin{equation}
        \begin{aligned}
        \mathbf{h}_1,\dots,\mathbf{h}_M
        &=
        \mathrm{TransformerEnc}\!\bigl(\\
        &\qquad
        \mathrm{PosEnc}(\mathbf{e}_1,\dots,\mathbf{e}_M)
        \bigr),
        \end{aligned}
        \end{equation}
        where $\mathrm{PosEnc}$ adds sinusoidal positional encodings to each input vector as in \citet{vaswani2017attention}.

        We then apply a linear classifier to each contextualized unit representation $\mathbf{h}_j$:
        \begin{equation}
            s_j=\mathbf{w}^{\top}\mathbf{h}_j+b.
        \end{equation}
        

        For each training program, we map the first buggy token annotated by \texttt{Collu-Bench} to its aligned lexical unit. Let $j^\star \in \{1,\dots,M\}$ denote the index of the buggy lexical unit in the AST-aligned sequence $\mathcal{T}(\mathcal{C})$. 
        The diagnostic model therefore learns to identify the first likely bug rather than to enumerate every bug in the program.
        We train the model with position-level cross-entropy over lexical-unit logits:
        \begin{equation}
            \mathcal{L}_{\text{FL}}
            =
            -\log
            \frac{\exp(s_{j^\star})}
            {\sum_{m=1}^{M}\exp(s_m)}.
            \label{eq:position_softmax}
        \end{equation}
        All trainable components of the diagnostic model are optimized end-to-end with $\mathcal{L}_{\text{FL}}$.

        At inference time, we treat each code line as a region composed of multiple lexical units, analogous to a region in an image, and apply max pooling over their suspiciousness scores. Specifically, for each line $c_i$ containing the lexical units $\mathcal{T}(c_i)$, we define its line-level suspiciousness score as
        \begin{equation}
            r_i=\max_{t_j \in \mathcal{T}(c_i)} s_j.
        \end{equation}
        We compare this max pooling design with an alternative design in Section~\ref{sec:alternativedesign}.

        The diagnostic model ranks lines in descending order of their suspiciousness scores, yielding $\mathcal{R}$. The refinement loop uses the top-$k$ lines in $\mathcal{R}$ for targeted refinement. 

\subsection{Top-$k$ Candidate Search}
\label{sec:bon}
    Given a code generation task and an initial program $\mathcal{C}^{(0)}$, {\method} iteratively refines the candidate program. At iteration $t$, {\method} executes the current program $\mathcal{C}^{(t)}$ against the public tests in a sandbox and obtains execution feedback $\xi^{(t)}$. 
    If the program passes all test cases, the loop terminates and returns the current program as the solution. 
    Otherwise, the diagnostic model produces a ranking of its suspicious lines, $\mathcal{R}^{(t)}$.

    Rather than relying only on the top-ranked line, {\method} searches over the top-$k$ suspicious lines. 
    For each selected line, {\method} constructs a targeted refinement prompt containing the code generation task, the current program, the execution feedback, and that suspicious line. 
    The language model generates one candidate revision for each prompt, producing $k$ candidate revisions at the current iteration.  
    Thus, $k=1$ uses only the most suspicious line, while larger $k$ values explore multiple plausible fault locations.

    We execute all candidate revisions against the public tests\footnote{During refinement, execution feedback and candidate selection use only public tests. Hidden tests are never used during refinement and are only used for final evaluation.}. If any candidate passes all tests, the loop terminates and returns a passing candidate as the final solution. 
    Otherwise, {\method} prefers the candidate that passes more tests or resolves more failures as $\mathcal{C}^{(t+1)}$.
    When candidates pass the same number of tests, we prefer the shortest candidate for conciseness. 
    Refinement continues until a candidate passes all public tests or the maximum iteration budget is reached. 
    Because the diagnostic model targets the first suspicious location, the language model generates a complete revised program at each step. Thus, a revision can modify multiple lines or insert missing code. {\method} then reruns execution and diagnosis, allowing later iterations to address any remaining errors.
\section{Experiments}
\label{sec:experiments}

\subsection{Experimental Setup}
    \paragraph{Benchmarks.}
    We evaluate {\method} on two representative code generation benchmarks. LiveCodeBench~\citep{jain2024livecodebench} continually incorporates
    recent problems from competitive programming platforms, including LeetCode, AtCoder, and Codeforces, which mitigates contamination risks.    BigCodeBench~\citep{zhuo2024bigcodebench} contains practical programming tasks that require diverse libraries and function calls. Neither benchmark is included in the source data used to construct the \texttt{Collu-Bench} training split.
    We report Pass@1, the percentage of tasks for which the final generated solution passes all test cases.
    
    \paragraph{Base LLMs.}
    We evaluate {\method} with five base LLMs spanning different model families and scales, including Qwen3-8B~\citep{yang2025qwen3}, Llama-3.3-70B~\citep{grattafiori2024llama}, GPT-4o-mini~\citep{hurst2024gpt}, GLM-5.1~\citep{glm5team2026glm5vibecodingagentic}, and DeepSeek-V4-Flash~\citep{deepseekai2026deepseekv4}. 
    
    \paragraph{Comparison Baselines.}
    We compare {\method} with three iterative refinement methods and two candidate-search baselines. First,  
    Self-Debugging \citep{chen2023teaching} generates an explanation of the current program and combines it with the execution results for refinement. 
    Self-Refine \citep{madaan2023selfrefine} prompts the model to generate self-feedback for code refinement. 
    NL-Debugging \citep{zhang2025nldebugging} translates the code solution into natural language and performs debugging in the natural language to guide code refinement. 
    Best-of-N \citep{snell2024scaling} generates 10 independent solutions and retains the one that passes the most public tests. We also equip NL-Debugging with a candidate-search budget of $k=10$. At each iteration, it generates 10 candidate revisions and, following the same selection rule as {\method}, retains the candidate that passes the most public tests.
    Following prior work~\citep{zhang2025nldebugging,chen2023teaching}, we also include the original LLM performance as a weak baseline. 
    
    \paragraph{Implementation Details.}
    The diagnostic model in {\method} is a bidirectional transformer encoder with 6 layers and 8 attention heads. 
    We train the full diagnostic model from scratch, including the input embeddings, transformer encoder, and linear prediction head. 
    Following \citet{zhang2025nldebugging,amini2025variational}, we use a maximum refinement iteration budget of $N$=5 and a candidate-search budget of $k$=10. 
    We also report the $k$=1 variant, which uses only the top-ranked suspicious region and removes expanded candidate search. Additional details are provided in Appendix~\ref{appendix: exp details}.

\begin{table*}[t]
\centering
\resizebox{0.90\textwidth}{!}{%
\begin{tabular}{@{}l|ccccc@{}}
\toprule
\multicolumn{6}{c}{\textbf{LiveCodeBench}} \\
\midrule
\textbf{Method} & \textbf{Qwen3-8B} & \textbf{Llama-3.3-70B} & \textbf{GPT-4o-mini} & \textbf{GLM-5.1} & \textbf{DeepSeek-V4-Flash} \\
\midrule
LLM alone & 30.86 & 46.86 & 44.00 & 54.86 & 58.86 \\
Self-Debugging & 38.29\gain{7.43} & 58.86\gain{12.00} & 51.43\gain{7.43} & 69.71\gain{14.85} & 75.43\gain{16.57} \\
Self-Refine & 36.00\gain{5.14} & 50.29\gain{3.43} & 48.00\gain{4.00} & 70.29\gain{15.43} & 68.00\gain{9.14} \\
NL-Debugging & \textbf{44.57}\gain{13.71} & 61.71\gain{14.85} & 53.14\gain{9.14} & 71.43\gain{16.57} & 76.57\gain{17.71} \\
\textbf{{\method}} ($k$=1) & 40.00\gain{9.14} & \textbf{64.57}\gain{17.71} & \textbf{54.86}\gain{10.86} & \textbf{81.14}\gain{26.28} & \textbf{81.71}\gain{22.85} \\
\midrule
Best-of-N ($N$=10) & 49.14\gain{18.28} & 70.29\gain{23.43} & 63.43\gain{19.43} & 77.71\gain{22.85} & 84.00\gain{25.14} \\
NL-Debugging ($k$=10) & \textbf{49.71}\gain{18.85} & 73.14\gain{26.28} & 61.71\gain{17.71} & 79.43\gain{24.57} & 85.14\gain{26.28} \\
\textbf{{\method}} ($k$=10) & {46.86}\gain{16.00} & \textbf{76.57}\gain{29.71} & \textbf{65.14}\gain{21.14} & \textbf{86.86}\gain{32.00} & \textbf{89.71}\gain{30.85} \\
\midrule
\multicolumn{6}{c}{\textbf{BigCodeBench}} \\
\midrule
LLM alone & 34.04 & 37.63 & 36.23 & 44.56 & 47.46 \\
Self-Debugging & 52.89\gain{18.85} & 56.58\gain{18.95} & 55.00\gain{18.77} & 68.51\gain{23.95} & 65.88\gain{18.42} \\
Self-Refine & 54.12\gain{20.08} & 57.46\gain{19.83} & 55.53\gain{19.30} & \textbf{71.75}\gain{27.19} & 68.68\gain{21.22} \\
NL-Debugging & 55.00\gain{20.96} & 58.33\gain{20.70} & 56.67\gain{20.44} & 69.82\gain{25.26} & 67.37\gain{19.91} \\
\textbf{{\method}} ($k$=1) & \textbf{57.63}\gain{23.59} & \textbf{59.39}\gain{21.76} & \textbf{58.51}\gain{22.28} & 71.23\gain{26.67} & \textbf{71.67}\gain{24.21} \\
\midrule
Best-of-N ($N$=10) & 48.95\gain{14.91} & 49.91\gain{12.28} & 62.02\gain{25.79} & 57.81\gain{13.25} & 62.63\gain{15.17} \\
NL-Debugging ($k$=10) & {61.93}\gain{27.89} & {71.49}\gain{33.86} & 64.91\gain{28.68} & {75.18}\gain{30.62} & 73.86\gain{26.40} \\
\textbf{{\method}} ($k$=10) & \textbf{64.47}\gain{30.43} & \textbf{72.81}\gain{35.18} & \textbf{69.65}\gain{33.42} & \textbf{78.42}\gain{33.86} & \textbf{79.91}\gain{32.45} \\
\bottomrule
\end{tabular}%
}
\caption{Pass@1 (\%) comparison between {\method} and iterative refinement baselines. For {\method}, $k$=1 uses a single candidate revision, whereas $k$=10 denotes the candidate-search setting. Arrows indicate gains or drops relative to the LLM alone. The best result under the same candidate-search setting is highlighted in bold.}
\label{tab:main_results}
\end{table*}

\subsection{Main Results}
    Table~\ref{tab:main_results} reports the end-to-end Pass@1 results. 
    Even without candidate search, {\method} ($k$=1) already improves on the iterative refinement baselines.
    It outperforms Self-Debugging in all settings and surpasses both Self-Refine and NL-Debugging in 9 of the 10 settings.
    Specifically, {\method} ($k$=1) achieves absolute Pass@1 gains of 1.72\% to 9.71\% over the strongest baseline, NL-Debugging, in four of five settings on LiveCodeBench. 
    On BigCodeBench, {\method} ($k$=1) provides consistent gains over NL-Debugging across all five models. 
    These results show that diagnostic localization improves iterative refinement even before expanded candidate search is applied. 

    Expanded candidate search allows {\method} to explore multiple plausible fault regions instead of committing to the highest-ranked one. 
    Compared with $k=1$, {\method} with $k=10$ improves Pass@1 by 8.97 points on average, with gains across all settings. The improvements range from 6.84 to 13.42 points on BigCodeBench and from 5.72 to 12.00 points on LiveCodeBench. 
    With these gains, {\method} ($k$=10) achieves the best performance in most settings, including both smaller and larger models as well as earlier and more recent ones. 

    To isolate the contribution of candidate search, we compare {\method} ($k$=10) with Best-of-N ($k=10$) and NL-Debugging ($k$=10). 
    Under the same candidate-search settings, {\method} outperforms Best-of-N by 4.11 points on LiveCodeBench and 16.79 points on BigCodeBench on average. It also outperforms NL-Debugging by 3.20 points on LiveCodeBench and 3.58 points on BigCodeBench. 
    

    Overall, these results show that precise diagnostic localization and candidate search play complementary roles. Localization identifies likely refinement regions, while expanded search mitigates uncertainty in the predicted fault location.


\begin{table}[ht]
    \centering
    \setlength{\tabcolsep}{3pt}
    \resizebox{0.95\linewidth}{!}{
    \scriptsize
    \begin{tabular}{l|cc}
    \toprule
    \textbf{Method} 
    & LCB & BCB \\
    \midrule
    Execution feedback only ($k$=1) & 50.29 & 53.77 \\
    Fine-grained feedback only ($k$=1) & 50.86 & 50.35 \\
    {\method} ($k$=1) & \textbf{54.86} & \textbf{58.51} \\
    \midrule
    Execution feedback only ($k$=10) &  58.29 & 59.47 \\
    Fine-grained feedback only ($k$=10) & 55.43 & 58.42 \\
    {\method} ($k$=10) & \textbf{65.14} & \textbf{69.65} \\
    \bottomrule
    \end{tabular}
    }
    \caption{Ablation results. ``Execution feedback only'' removes fine-grained diagnostic feedback, while ``Fine-grained feedback only'' removes execution feedback.}
    \label{tab:ablation}
\end{table}

\subsection{Ablation Study}
    \label{sec:ablations}
    The main results already show that ablating the search procedure by setting $k$ to 1 outperforms other single-candidate baselines. 
    In this section, we therefore focus the ablation study on the contribution of each feedback source. 
    Table~\ref{tab:ablation} isolates the contributions of diagnostic and execution feedback using GPT-4o-mini on LiveCodeBench and BigCodeBench. We compare full {\method} with variants that remove one feedback source from the refinement prompt at both $k=1$ and $k=10$.

    \paragraph{Fine-Grained Feedback.}
    Fine-grained diagnostic feedback improves refinement under both search budgets. The execution-only variant retains execution feedback in the refinement prompt but removes the predicted suspicious line. 
    At $k$=1, execution-only refinement reaches 50.29\% Pass@1 on LiveCodeBench and 53.77\% on BigCodeBench, while full {\method} reaches 54.86\% and 58.51\%. 
    At $k$=10, execution-only refinement reaches 58.29\% and 59.47\%, while full {\method} reaches 65.14\% and 69.65\%. 
    The consistent gaps show that fine-grained feedback provides useful localization guidance on top of execution feedback in both the single-candidate and expanded-search settings.

    \paragraph{Execution Feedback.}
    Execution feedback also improves refinement under both search budgets. The diagnostic-only variant includes the predicted suspicious line but omits execution feedback. 
    At $k$=1, diagnostic-only refinement reaches 50.86\% Pass@1 on LiveCodeBench and 50.35\% on BigCodeBench, below the full {\method} scores of 54.86\% and 58.51\%. 
    At $k$=10, diagnostic-only refinement reaches 55.43\% and 58.42\%, while full {\method} reaches 65.14\% and 69.65\%. 
    Together, the two ablations show that the signals play complementary roles: fine-grained feedback identifies where to revise, while execution feedback describes the observed failure.

\begin{figure*}[htbp]
    \centering
    \begin{minipage}[b]{0.48\textwidth}
        \centering
        \includegraphics[width=\linewidth]{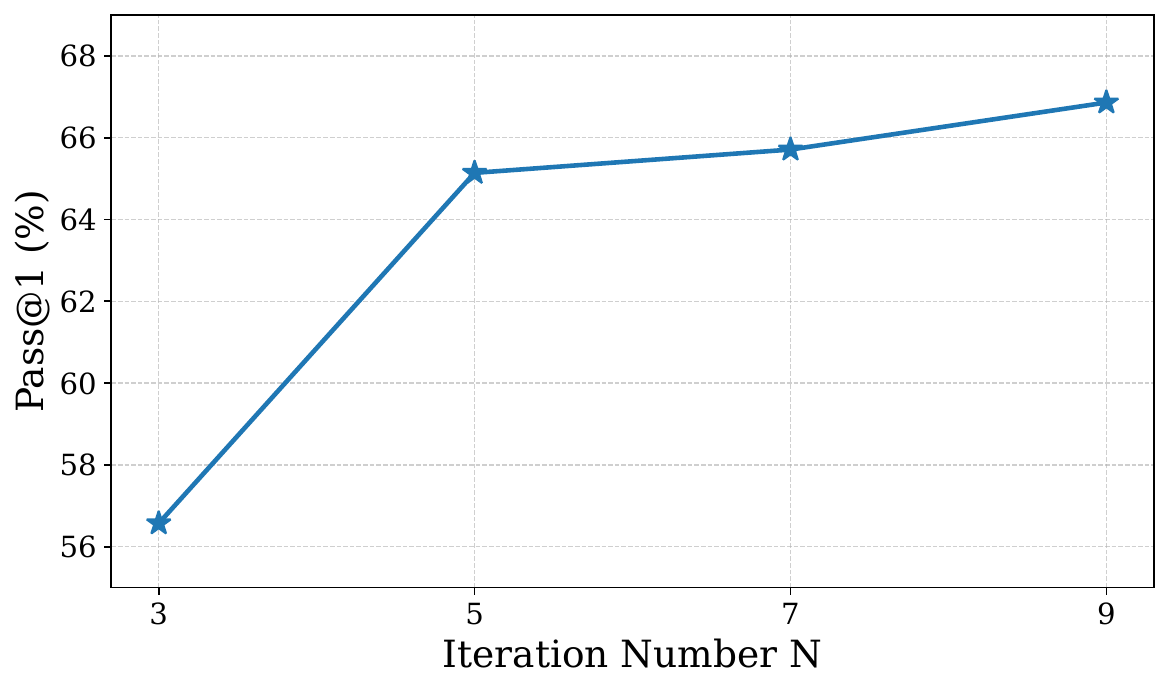}
        \label{fig:varying_n}
    \end{minipage}
    \hfill
    \begin{minipage}[b]{0.48\textwidth}
        \centering
        \includegraphics[width=\linewidth]{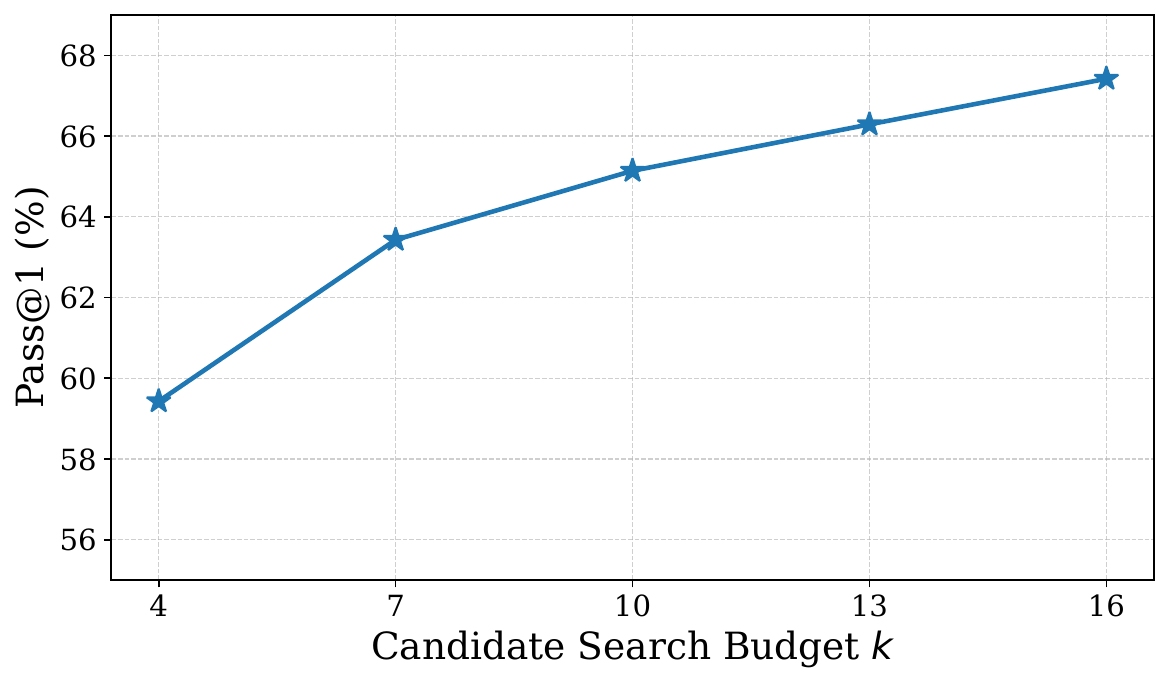}
        \label{fig:varying_k}
    \end{minipage}
    \caption{Effect of the iteration budget $N$ and candidate-search budget $k$ on GPT-4o-mini on LiveCodeBench.}
    \label{fig:effect_search_budget}
\end{figure*}

\subsection{Alternative Aggregation Design}
    \label{sec:alternativedesign}
    The diagnostic model predicts suspiciousness scores for lexical units, while the refinement prompt uses line-level feedback. 
    We therefore aggregate the lexical-unit scores within each line. To evaluate this aggregation step, we compare the max pooling used in {\method} with average pooling. The comparison uses GPT-4o-mini with the default configuration. Average pooling computes the mean logit over the lexical units on each line:
    
    \begin{equation}
        r_i
        =
        \frac{1}{|\mathcal{T}(c_i)|}
        \sum_{t_j\in\mathcal{T}(c_i)} s_j.
    \end{equation}

\begin{table}[ht]
    \centering
    \setlength{\tabcolsep}{4pt}
    \resizebox{0.82\linewidth}{!}{
    \scriptsize
    \begin{tabular}{l|cc}
    \toprule
    \textbf{Method} & LCB & BCB \\
    \midrule
    {\method} (Average pooling) & 59.43 & 62.54 \\
    {\method} (Max pooling) & \textbf{65.14} & \textbf{69.65} \\
    \bottomrule
    \end{tabular}
    }
    \caption{Alternative line-level aggregation designs.}
    \label{tab:aggregation_design}
\end{table}

    As shown in Table~\ref{tab:aggregation_design}, max pooling raises Pass@1 from 59.43\% to 65.14\% on LiveCodeBench and from 62.54\% to 69.65\% on BigCodeBench. This is because max pooling retains a strong suspiciousness signal from a single faulty unit, whereas averaging can dilute it with many non-suspicious units on the same line.

\subsection{Diagnostic Model Performance}
    \label{sec:diagnostic}
    We evaluate the diagnostic model on 100 randomly selected failed programs from LiveCodeBench. Following the Collu-Bench workflow, we manually annotated the reference fault locations by inspecting the task description, generated program, and observed failure, and then marked the first line containing the faulty logic.\footnote{Annotation and baseline configurations are detailed in Appendix~\ref{app:fl_details}}
    We report Top-1, Top-3, and Top-10 localization accuracy. 
    Direct prompting asks an LLM to identify suspicious lines from the code. 
    We also compare with FlexFL~\citep{flexfl}, which refines conventional fault-localization results with an LLM, and LLMAO~\citep{yang2024large} and BAP~\citep{stein2025where}, which use model-internal signals for fault localization. 

\begin{table}[ht]
    \centering
    \setlength{\tabcolsep}{4pt}
    \resizebox{0.82\linewidth}{!}{
    \scriptsize
    \begin{tabular}{l|ccc}
    \toprule
    \textbf{Method} & Top 1 & Top 3 & Top 10 \\
    \midrule
    Direct Prompting & 51 & 70 & 79 \\
    FlexFL & 38 & 55 & 76 \\
    LLMAO & 63 & 74 & 87 \\
    BAP & 54 & 62 & 84 \\
    \midrule
    {\method} & \textbf{67} & \textbf{75} & \textbf{89} \\
    \bottomrule
    \end{tabular}
    }
    \caption{Performance comparison of the diagnostic model in {\method} with fault localization baselines on LiveCodeBench.}
    \label{tab:fl_comparison}
\end{table}

    As shown in Table~\ref{tab:fl_comparison}, {\method} achieves the strongest localization performance across all settings. 
    Specifically, it outperforms the strongest baseline, LLMAO, with an absolute improvement of 4\% in Top-1 accuracy. 
    
\subsection{Effect of Search Budget}
    \label{sec:search_budget}
    We analyze the effect of two search-budget parameters: the maximum number of refinement iterations $N$ and the number of suspicious lines $k$ explored at each iteration. We vary one budget at a time while holding the other fixed on LiveCodeBench with GPT-4o-mini. Figure~\ref{fig:effect_search_budget} shows the results. 
    
    Increasing the number of iterations from 3 to 5 improves Pass@1 from 56.57\% to 65.14\%. 
    Further increasing the budget yields smaller gains, reaching 65.71\% at 7 iterations and 66.86\% at 9 iterations. 
    Similarly, increasing the candidate budget $k$ from 4 to 10 improves Pass@1 from 59.43\% to 65.14\%, while increasing $k$ beyond 10 yields smaller returns relative to the added refinement cost, reaching 66.29\% at $k=13$ and 67.43\% at $k=16$. 
    The default setting ($N=5$ and $k=10$) provides the best balance between computational cost and refinement performance.

\subsection{Efficiency Analysis}
    To measure computational cost, we report the average runtime and total token consumption per task on LiveCodeBench in Table~\ref{tab:efficiency}. 
    When $k$ is set to 1, {\method} takes 73.76 seconds and consumes 4932.09 tokens per task. In comparison, Self-Debugging and Self-Refine take 95.96 and 91.94 seconds and consume 4850.64 and 6571.56 tokens, respectively.
    Thus, {\method} has the lowest runtime, while its token consumption is comparable to Self-Debugging and lower than Self-Refine. This runtime advantage arises because {\method} uses a lightweight diagnostic model for fine-grained fault localization and therefore does not require an additional LLM pass to generate feedback.
    NL-Debugging adds a natural-language debugging step before code revision. It is therefore the most expensive of the three baselines without parallel sampling, taking 211.68 seconds and consuming 16511.29 tokens per task. 
    
    Searching over 10 candidate refinements ($k$=10) increases the runtime cost to 610.19 seconds, and token consumption to 28415.10 tokens per task. This cost is slightly higher than Best-of-N ($N=10$), which takes 536.81 seconds and consumes 25475.66 tokens, but lower than NL-Debugging ($k=10$), which takes 971.36 seconds and consumes 49134.73 tokens.
    Although expanding search from $k=1$ to $k=10$ increases the runtime and token cost, it improves Pass@1 by 10.28 points on LiveCodeBench and 11.14 points on BigCodeBench for GPT-4o-mini. The added inference cost therefore accompanies substantial performance gains.

\begin{table}[ht]
    \centering
    \setlength{\tabcolsep}{2pt}
    \resizebox{0.9\linewidth}{!}{
    \scriptsize
    \begin{tabular}{l|cc}
    \toprule
    \textbf{Method} & Seconds/task & \# Tokens/task\\
    \midrule
    Self-Debugging & 95.96 & 4850.64 \\ 
    Self-Refine & 91.94 & 6571.56 \\ 
    NL-Debugging & 211.68 & 16511.29 \\
    Best-of-N ($N$=10) & 536.81 & 25475.66 \\
    NL-Debugging ($k$=10) & 971.36 & 49134.73 \\
    \midrule
    {\method} ($k$=1) & 73.76 & 4932.09 \\
    {\method} ($k$=10) & 610.19 & 28415.10 \\
    \bottomrule
    \end{tabular}
    }
    \caption{Runtime and token cost comparison on LiveCodeBench.}
    \label{tab:efficiency}
\end{table}

\subsection{Failure Case Analysis}
\label{sec:analysis}

To understand the limitations of {\method}, we manually inspected 418 final solutions that remain incorrect on BigCodeBench and LiveCodeBench and grouped their failures into three categories. Appendix~\ref{sec:appendix_failure_cases} provides an example of each category.

\paragraph{Language Model Reasoning Failures (253/418).}
    Most failures occur when localization identifies useful regions, but the language model still lacks the capability to perform the required reasoning or algorithmic change. 
    For example, in the purchasing problem in Table~\ref{tab:failure_case_pattern1}, where buying $k$ units of product $i$ costs $k^2P_i$, the model incorrectly uses binary search on the target number of units and validates it with a naive loop. 
    A correct solution instead needs to reason about marginal costs and purchase units in increasing marginal cost order. 
    In such cases, the bottleneck is not only where to edit the program, but also whether the language model can select the right problem-solving strategy. 

\paragraph{Misinterpretation of Task Requirements (88/418).}
    Some failures stem from an incorrect understanding of the problem specification. 
    The model may implement a plausible algorithmic component but apply the wrong constraint, formula, or objective in the final computation. 
    For example, in the stone merging problem in Table~\ref{tab:failure_case_pattern2}, the model applies a linear-basis solution for subset XOR, but the task requires reasoning about XOR values after whole-bag merge operations. 
    These errors are difficult to refine through localization alone because the faulty code often reflects a mistaken interpretation of the task rather than a localized implementation bug.

\paragraph{Long-Range State and Simulation Errors (77/418).}
    The remaining failures often involve complex state transitions, multi-stage simulations, or long-range variable dependencies. 
    In these cases, the model may fix one local transition while breaking a later dependency. 
    For example, in the simulation problem in Table~\ref{tab:failure_case_pattern3}, the model represents the history of which adults still have stones with a single global counter, discarding information needed for later years. 
    This suggests that future work may need feedback from intermediate program states in addition to line-level localization.

\section{Related Work}
\label{sec:related_work}

\vspace{-0.1cm}
\paragraph{Iterative Code Refinement.}
Iterative code refinement typically follows a refinement loop in which a model generates code, receives feedback, and then revises the generated code. 
Prior work has explored different forms of feedback for this process. 
Self-Debugging~\citep{chen2023teaching}, Self-Edit~\citep{zhang-etal-2023-self}, Cycle~\citep{ding2024cycle}, and OpenCodeInterpreter~\cite{zheng2024opencodeinterpreter} use execution feedback to guide code revision. By contrast, Self-Refine~\citep{madaan2023selfrefine} relies on the model's own critique to improve its output. \citet{welleck2022generating} studies self-correction with model-generated feedback. More recently, NL-Debugging translates the code solution into natural language and performs debugging in the natural language to guide code refinement~\citep{zhang2025nldebugging}. However, such signals are either too coarse-grained or too high-level to tell the model where the code should be revised.
{\method} addresses this limitation by augmenting iterative refinement with line-level diagnostic feedback.

\paragraph{Fault Localization.}
Traditional fault localization (FL) methods often use test coverage information to predict the fault location. For instance,  
spectrum-based FL analyzes runtime spectra from passing and failing test cases and ranks program entities with suspiciousness metrics such as Jaccard~\citep{abreu2006evaluation}, Ochiai~\citep{abreu2006evaluation}, and DStar~\citep{wong2013dstar}. Learning-based FL extends this approach by training ranking or prediction models on execution spectra and other program features, such as code complexity metrics and error messages~\citep{li2019deepfl,sohn2017fluccs}. 

More recently, LLM-based FL methods have used language models to reason about suspicious code regions. 
AgentFL~\citep{qin2024agentfl} prompts LLMs with diversified debugging information from failed tests. 
FlexFL~\citep{flexfl} decomposes localization into two stages that use LLMs to refine localization results obtained from one or more existing FL techniques. 
SOAPFL~\citep{10891926} further standardizes LLM-based localization into structured reasoning stages, including comprehension, navigation, and confirmation. 
Another line of work probes model-internal signals to predict buggy regions. 
LLMAO~\citep{yang2024large} fine-tunes bidirectional adapters to predict line-level bugginess, while 
BAP~\citep{stein2025where} uses weak supervision to interpret attention patterns for scalable localization.

Compared with these FL methods, {\method} trains a lightweight diagnostic model to localize suspicious code regions from the language model's probabilistic signals. 
Our evaluation shows that this diagnostic model outperforms recent methods including FlexFL~\citep{flexfl}, LLMAO~\citep{yang2024large}, and BAP~\citep{stein2025where}.
\section{Conclusion}
We presented {\method}, an iterative framework with a lightweight diagnostic model that predicts line-level suspiciousness signals for bug localization and refinement. 
To account for the uncertainty of diagnostic predictions, {\method} searches over the top-$k$ suspicious lines and selects candidate revisions using execution outcomes. 
Experiments on LiveCodeBench and BigCodeBench with five base LLMs show that {\method} outperforms existing iterative refinement methods in most settings, even without candidate search. 
Ablations confirm that fine-grained diagnostic feedback and candidate search contribute significantly to overall gains.
These results suggest that fine-grained localization signals serve as reliable guidance for code refinement.

\section*{Limitations}
Our framework has three main limitations. 
First, the iterative refinement process introduces overhead. 
In our efficiency analysis, we show that this overhead is moderate compared with other iterative refinement methods. 
Nevertheless, we acknowledge that the additional cost may become non-negligible when the number of refinement rounds or the top-$k$ candidate budget is large.

Second, as discussed in our failure-case analysis, the capability of the language model affects the performance. 
For example, failures arise from an incorrect understanding of the problem specification. 
These errors are difficult to resolve through localization alone because the faulty code often reflects a mistaken interpretation of the task rather than a localized implementation bug.

Third, {\method} uses token probabilities produced during code generation as input to the diagnostic model. These internal signals are used to predict the bug location. {\method} can therefore be applied to any model that exposes these internal signals, including all models evaluated in this paper. However, some commercial APIs do not expose these signals, so the method cannot be applied to them currently.

\bibliography{custom}

\newpage
\appendix
\section{Implementation Details}
\label{appendix:exp_details}

\subsection{Model Architecture}
\label{app:model}

\paragraph{Embeddings and projection.}
We use three learnable modules to embed each lexical unit:
(i) a token embedding layer $\mathrm{Emb}_{tok}$,
(ii) a type embedding layer $\mathrm{Emb}_{type}$,
and (iii) a linear projection $\mathrm{Proj}$ for the scalar confidence value.
In our implementation, the token embedding dimension is $256$, the syntactic-type embedding dimension is $128$, and the probability projection dimension is $128$. The concatenated vector has dimension $512$.

\subsection{Training Setup}
\label{app:training}

We optimize the diagnostic model using AdamW with a learning rate of $10^{-4}$ for $15$ epochs and a batch size of $32$.
Given the ground-truth buggy unit index $j^\star$, we minimize a position-level cross-entropy loss over the logits $\{s_j\}_{j=1}^{M}$.
We select the best checkpoint according to validation accuracy (exact match of the predicted index).

\subsection{Experimental Details}
    \label{appendix: exp details}
    We trained our diagnostic model on a single NVIDIA GeForce RTX 4090 GPU. For training and evaluating the diagnostic model on Collu-Bench, the dataset contains 13130 samples in total, split into 10504 training samples and 2626 test samples. We train a single diagnostic model on this training split and share it across all five base LLMs in our end-to-end experiments without per-model retraining. Because \texttt{Collu-Bench} contains programs and token probabilities from 11 different generator LLMs, the training data does not tie the diagnostic model to any single generator. The model was trained for 15 epochs, with a total training time of approximately 10 minutes under this hardware configuration. 

    As for evaluation benchmarks, we use the \texttt{code\_generation\_lite} subset (\texttt{release\_v6}) of LiveCodeBench, which contains 175 tasks, and the full BigCodeBench suite, which contains 1140 tasks.

\subsection{Fault-Localization Evaluation Details}
    \label{app:fl_details}
    
    We manually annotated the 100 randomly selected failed LiveCodeBench programs used in Section~\ref{sec:diagnostic}. For each program, we inspected the task, the generated code, and the observed failure, and marked the first line containing the faulty logic as the reference fault location.
    
    We used official parameter configurations for the fault-localization baselines whenever they were available. For LLMAO and BAP, we used Llama~3.2~11B as the backbone. We constructed their fine-tuning examples from the same \texttt{Collu-Bench} training split used for {\method}, following each method's original data-construction procedure. Their fine-tuning configurations follow the corresponding papers and official repositories. For FlexFL, we used its released prompt with GPT-4o on the same 100 manually annotated programs.

\subsection{Annotation Details}
\label{app:annotation_details}

For the failure case analysis in Section~\ref{sec:analysis}, one author inspected all 418 incorrect final solutions on BigCodeBench and LiveCodeBench. For each solution, the annotator read the task description, the final program, and the execution results, and assigned the solution to one of the three failure categories. The three categories contain 253, 88, and 77 solutions, corresponding to the reported 60.53\%, 21.05\%, and 18.42\%.

For the fault-localization evaluation in Section~\ref{sec:diagnostic}, one author labeled the reference fault line of each of the 100 sampled LiveCodeBench programs, following the procedure described in Appendix~\ref{app:fl_details}. The annotator did not have access to the predictions of {\method} or any baseline during labeling. We release the 100 annotated programs and their fault labels together with our code.

\subsection{Direct Prompting}
    \label{appendix:directprompting}

    We include the prompt in Table~\ref{tab:prompt} for our direct prompting baseline, where the model is asked to directly identify and rank the most suspicious lines in the given code. The input code is formatted with explicit line numbers. 
    \texttt{<LINE\_NUMBERED\_CODE>} is the full program text where each line is prefixed with its 1-indexed line number (e.g., \texttt{1: ...}, \texttt{2: ...}).

    \begin{table}[ht]
          \centering
          \small
          \begin{tabular}{>{\raggedright\arraybackslash\ttfamily}p{0.45\textwidth}<{}}
              \toprule

\headercolorlong
\textbf{System message}\\
You are a helpful code analysis assistant.\\\\
\headercolorlong
\textbf{Direct Prompting}\\
You are an expert code reviewer. Your task is to identify lines that may contain hallucinations (factual errors, non-existent APIs, incorrect function usage, or fabricated information) in the following code.\\\\
The code with line numbers:
\begin{verbatim}
<LINE_NUMBERED_CODE>
\end{verbatim}

Instructions:
\begin{enumerate}
\item Analyze the code carefully
\item Identify the lines that most likely contain hallucinations
\item Rank them from most suspicious to least suspicious
\item Return 10 line numbers, ordered by suspicion level (most suspicious first)
\end{enumerate}

Your response should be in the following format (one line number per line, most suspicious first):
\begin{verbatim}
1. Line <number>: <brief reason>
2. Line <number>: <brief reason>
...
\end{verbatim}

% ...
\\
\bottomrule
          \end{tabular}
          \caption{Direct Prompting LLM for fault localization.}
          \label{tab:prompt}
        \end{table}

\begin{table}[ht]
    \centering
    \setlength{\tabcolsep}{24pt}
    \resizebox{0.95\linewidth}{!}{
    \scriptsize
    \begin{tabular}{l r}
    \hline
    Type & ID \\
    \hline
    \texttt{<PAD>} & 0 \\
    \texttt{<UNK>} & 1 \\
    \texttt{Constant} & 2 \\
    \texttt{Delimiter} & 3 \\
    \texttt{EOS} & 4 \\
    \texttt{Error} & 5 \\
    \texttt{Identifier} & 6 \\
    \texttt{Keyword} & 7 \\
    \texttt{Operator} & 8 \\
    \texttt{Space} & 9 \\
    \texttt{Type} & 10 \\
    \hline
    \end{tabular}
    }
    \caption{Syntactic type vocabulary used by the diagnostic model.}
    \label{tab:type-inventory}
    \vspace{-0.4cm}
\end{table}

\lstset{
  basicstyle=\ttfamily\scriptsize,
  breaklines=true,
  columns=fullflexible,
  keepspaces=true,
  showstringspaces=false,
  aboveskip=0pt,
  belowskip=0pt
}

\section{Additional Experiment Results}
\subsection{Diagnostic Model Performance}
\label{app:collu_test}

Section~\ref{sec:diagnostic} evaluates the diagnostic model on 100 LiveCodeBench programs with fault locations that we annotated ourselves. In this section, we additionally compare all methods on the subset of held-out \texttt{Collu-Bench} test split: 300 randomly sampled buggy programs whose fault annotations come with the benchmark rather than from our own labeling. The baselines and configurations are identical to those in Section~\ref{sec:diagnostic}, and Table~\ref{tab:appfl_comparison_collu} reports Top-1, Top-3, and Top-10 localization accuracy (\%). {\method} performs best on all three metrics and outperforms the strongest baseline, LLMAO, by 4.33 points in Top-1 accuracy. The absolute accuracy is higher than on LiveCodeBench because the test split follows the training distribution of the learned methods.

\begin{table}[ht]
    \centering
    \setlength{\tabcolsep}{4pt}
    \resizebox{0.82\linewidth}{!}{
    \scriptsize
    \begin{tabular}{l|ccc}
    \toprule
    \textbf{Method} & Top 1 & Top 3 & Top 10 \\
    \midrule
    Direct Prompting & 59.67 & 88.67 & 94.67 \\ 
    FlexFL & 40.33 & 65.00 & 93.67 \\ 
    LLMAO & 89.00 & 95.67 & 98.67 \\ 
    BAP & 61.67 & 84.67  & 98.00 \\ 
    \midrule
    {\method} & \textbf{93.33} & \textbf{96.33} & \textbf{99.67} \\ 
    \bottomrule
    \end{tabular}
    }
    \caption{Performance comparison of the diagnostic module in {\method} with fault localization baselines on Collu-Bench Test Split.}
    \label{tab:appfl_comparison_collu}
    \vspace{-0.1cm}
\end{table}

\subsection{Length-Aware Confidence Aggregation}
\label{app:length_adjustment}

Equation~\ref{eq:lexical_confidence} defines the confidence of lexical unit $t_j$ as $p'_j=\prod_{n\in\mathcal{S}(j)}p_n$, where $\mathcal{S}(j)$ is the set of BPE-token indices aligned with $t_j$. We write this feature in the equivalent log-score space, $\ell'_j=\log p'_j=\sum_{n\in\mathcal{S}(j)}\log p_n$, where $n_j=|\mathcal{S}(j)|$. 
Because the product is affected by the number of aligned BPE tokens, we compare the original score with mean normalization and additive length compensation to remove this length effect.

\begin{table}[ht]
    \centering
    \setlength{\tabcolsep}{2pt}
    \resizebox{0.95\linewidth}{!}{
    \scriptsize
    \begin{tabular}{lcc}
    \toprule
    \textbf{Adjustment} & {\textbf{Log-space feature}} & {\textbf{Validation accuracy}} \\
    \midrule
    {None} & {$\ell'_j$} & {\textbf{0.7784}} \\
    {Mean} & {$\ell'_j/n_j$} & {0.7708} \\
    {Additive} & {$\ell'_j+\alpha(n_j-1)$} & {0.7586} \\
    \bottomrule
    \end{tabular}
    }
    \caption{{Validation accuracy under alternative lexical-unit confidence aggregations.}}
    \label{tab:length_adjustment}
    \vspace{-0.4cm}
\end{table}

{The mean design, $\ell'_j/n_j$, uses the average token log-probability, corresponding to standard length normalization in sequence scoring~\citep{murray-chiang-2018-correcting}. The additive variant adds a fixed reward for each extra BPE token, $\alpha(n_j-1)$, following the per-word reward approach~\citet{murray-chiang-2018-correcting}. We set $\alpha=0.2$, following the initialization used in that work.

All three aggregations results are close. The largest gap is only 1.98 validation points, and the original aggregation feature achieves the best performance. One likely reason is that AST lexical alignment confines each aggregation to a syntactic lexical unit, which limits the range of subword lengths under comparison. Another is that the aggregate confidence is only one scalar input; the model uses lexical identity, syntactic type, position, and full-program context as inputs.

\subsection{Efficiency}
Breaking down the runtime in Section \ref{tab:efficiency}, language model inference accounts for approximately 87\% of this runtime cost, while test execution accounts for most of the remaining cost. The lightweight diagnostic model contributes less than 1\% of the total runtime. Thus, most of the additional cost comes from generation and validation, reflecting the intended trade-off between computation and refinement quality. 

\section{Case Study}
\label{appendix:case}

To understand how {\method} refines programs, we present an example from LiveCodeBench. 
The problem gives $N$ foods and a calorie budget $X$. Each food has a vitamin type, a vitamin amount, and a calorie cost. The objective is to maximize the smallest total amount among the three vitamins. Thus, a correct solution must use one shared food subset that both stays within budget and balances the three vitamin totals. We provide complete examples in Table~\ref{tab:case-study-livecodebench} and Table~\ref{tab:case-study-failed-debugging}.

The initial solution incorrectly defines the dynamic programming (DP) state as \texttt{dp = [[0] * (X + 1) for \_ in range(4)]}, using rows 1 to 3 for the three vitamin types and optimizes each vitamin separately. 
This design treats each row as an independent knapsack table, which is the core mistake. 
Each vitamin row can spend the full calorie budget on a different subset of foods, so \texttt{dp[1][X]}, \texttt{dp[2][X]}, and \texttt{dp[3][X]} may not come from the same food selection. 
However, the task requires one shared subset whose total calories are within $X$. 
The final \texttt{min} computation therefore treats three separately optimized values as if they were achieved by the same selection, so the reported answer can be too large.

Execution feedback can show that the output is wrong, but it does not identify that the abstraction of the DP state is the root cause. 
{\method}'s diagnostic model ranks the DP initialization line as the most suspicious. This feedback points directly to the flawed state abstraction.

Guided by this diagnostic feedback, {\method} produces a refined solution with a different strategy. 
This solution uses binary search over a target value $m$ and checks whether one subset of foods can make all three vitamin totals at least $m$ within budget $X$. 
This formulation enforces the shared budget constraint and avoids the independent-DP mistake. 

By contrast, Self-Debugging keeps the three independent DP tables and adds a binary-search check over a single calorie value. Self-Refine only makes a minor revision, preserving the original per-vitamin DP tables and the final \texttt{min} over independently optimized values. 
NL-Debugging rewrites the table as \texttt{dp[c][v]}, but each entry still records only the best value for one vitamin rather than a joint food selection. 
All revisions continue to combine incompatible choices across vitamin types, reflecting a common refinement failure mode in which the model applies superficial patches without correcting the underlying abstraction. 
By localizing the suspicious DP state, {\method} helps refinement to address the root cause and produce a correct solution strategy.

\section{Syntactic Type}
\label{app:type-inventory}
Table~\ref{tab:type-inventory} lists the syntactic type vocabulary used by the diagnostic model. 
The vocabulary includes standard code categories, such as \texttt{Identifier}, \texttt{Keyword}, \texttt{Operator}, \texttt{Delimiter}, and \texttt{Constant}, together with preprocessing symbols. 
\texttt{<PAD>} marks padded positions, \texttt{<UNK>} marks unseen or unrecognized types, and \texttt{EOS}, \texttt{Space}, and \texttt{Error} are retained when they appear in the lexical-unit sequence.

\section{Prompt Templates}
\label{app:repair_prompts}

\lstdefinestyle{promptstyle}{
basicstyle=\ttfamily\small,
breaklines=true,
breakatwhitespace=false,
columns=fullflexible,
frame=single,
backgroundcolor=\color{gray!5},
rulecolor=\color{gray!40},
xleftmargin=1em,
xrightmargin=1em,
aboveskip=0.8em,
belowskip=0.8em
}

This section lists the prompt templates used for initial generation, refinement, ablations, and top-$k$ candidate search. 
Templates use placeholders for task text, failed code, execution feedback, and diagnostic feedback.

\subsection{Initial Code Generation Prompt}
\label{app:initial_generation_prompt}

\begin{table}[ht]
      \centering
      \begin{tabular}{>{\raggedright\arraybackslash\ttfamily\scriptsize}p{0.46\textwidth}<{}}
          \toprule

\headercolorlong
\textbf{Initial Code Generation Prompt}\\
Please complete the following Python code based on the problem description
and the provided function/class structure. Only provide the complete,
executable code block required to solve the problem.\\
<STARTER\_CODE>\\

\bottomrule
      \end{tabular}
      \caption{Initial code generation prompt.}
      \label{tab:initial_generation_prompt}
\end{table}

The initial prompt in Table~\ref{tab:initial_generation_prompt} asks the model to produce a complete executable solution from the benchmark-provided task scaffold.

Here, \texttt{<STARTER\_CODE>} denotes the benchmark-provided starter code with the problem description. 

\subsection{Iterative Refinement Prompt}
\label{app:general_repair_prompt}

For each failed attempt, the refinement prompt in Table~\ref{tab:general_refinement_prompt} includes the original task, the failed code, execution feedback, and optionally a suspicious line predicted by the diagnostic model.

\begin{table}[t]
      \centering
      \begin{tabular}{>{\raggedright\arraybackslash\ttfamily\scriptsize}p{0.46\textwidth}<{}}
          \toprule

\headercolorlong
\textbf{Iterative Refinement Prompt}\\
Original Task:\\
The original task was to complete the following Python function:\\
<INITIAL\_PROMPT\_OR\_TASK\_SCAFFOLD>\\
Previous Attempt Analysis:\\
Your last code attempt failed. Here is a detailed analysis of the failure.\\
Failure Context:\\
Input:\\
<FAILED\_TEST\_INPUT>\\
Expected Output:\\
<EXPECTED\_OUTPUT>\\
Your Actual Output:\\
<ACTUAL\_OUTPUT>\\
Execution Error or Traceback:\\
When tested, your code produced the following error:\\
<EXECUTION\_ERROR\_OR\_TRACEBACK>\\
Fine-grained Diagnostic Feedback:\\
The diagnostic model ranks the following line as suspicious:\\
Suspicious Line:\\
<SUSPICIOUS\_CODE\_LINE>\\
This line may be relevant to the observed execution failure.\\
Your Failed Code:\\
<PREVIOUS\_FAILED\_CODE>\\
Please provide a new, complete, and corrected version of the function
that resolves all identified issues.\\

\bottomrule
      \end{tabular}
      \caption{Iterative refinement prompt. The fine-grained diagnostic feedback block is included only when diagnostic feedback is enabled.}
      \label{tab:general_refinement_prompt}
\end{table}

\texttt{<EXECUTION\_ERROR\_OR\_TRACEBACK>} contains the observed assertion failure, runtime exception, timeout, or other execution result. Test input and output placeholders are filled when available.

\subsection{Execution Feedback Only Refinement Prompt}
\label{app:execution_repair_prompt}

The prompt in Table~\ref{tab:execution_only_refinement_prompt} is used in the ablation study that removes diagnostic feedback and keeps only execution feedback.

\begin{table}[t]
      \centering
      \begin{tabular}{>{\raggedright\arraybackslash\ttfamily\scriptsize}p{0.46\textwidth}<{}}
          \toprule

\headercolorlong
\textbf{Execution feedback only Refinement Prompt}\\
Original Task:\\
The original task was to complete the following Python function:\\
<INITIAL\_PROMPT\_OR\_TASK\_SCAFFOLD>\\
Previous Attempt Analysis:\\
Your last code attempt failed. Here is a detailed analysis of the failure.\\
Failure Context:\\
Input:\\
<FAILED\_TEST\_INPUT>\\
Expected Output:\\
<EXPECTED\_OUTPUT>\\
Your Actual Output:\\
<ACTUAL\_OUTPUT>\\
Execution Error or Traceback:\\
When tested, your code produced the following error:\\
<EXECUTION\_ERROR\_OR\_TRACEBACK>\\
Your Failed Code:\\
<PREVIOUS\_FAILED\_CODE>\\
Please provide a new, complete, and corrected version of the function
that resolves all identified issues.\\

\bottomrule
      \end{tabular}
      \caption{Execution feedback only refinement prompt.}
      \label{tab:execution_only_refinement_prompt}
\end{table}

\subsection{Fine-grained feedback only Refinement Prompt}
\label{app:diagnostic_only_repair_prompt}

The prompt in Table~\ref{tab:diagnostic_only_refinement_prompt} is used in the ablation study that removes detailed execution feedback and keeps only the diagnostic suspicious-line signal.

\begin{table}[t]
      \centering
      \begin{tabular}{>{\raggedright\arraybackslash\ttfamily\scriptsize}p{0.46\textwidth}<{}}
          \toprule

\headercolorlong
\textbf{Fine-grained feedback only Refinement Prompt}\\
Original Task:\\
The original task was to complete the following Python function:\\
<INITIAL\_PROMPT\_OR\_TASK\_SCAFFOLD>\\
Previous Attempt Analysis:\\
Your last code attempt failed one or more test cases.\\
Fine-grained Diagnostic Feedback:\\
The diagnostic model ranks the following line as suspicious:\\
Suspicious Line:\\
<SUSPICIOUS\_CODE\_LINE>\\
This line may be relevant to the observed failure.\\
Your Failed Code:\\
<PREVIOUS\_FAILED\_CODE>\\
Please provide a new, complete, and corrected version of the function
that resolves all identified issues.\\

\bottomrule
      \end{tabular}
      \caption{Fine-grained feedback only refinement prompt.}
      \label{tab:diagnostic_only_refinement_prompt}
\end{table}

\subsection{top-$k$ Candidate Search Prompt}
\label{app:top_k_prompt}

For top-$k$ candidate search, each branch uses the same diagnostic-guided refinement prompt in Table~\ref{tab:topk_candidate_search_prompt} with a different suspicious line from the top-$k$ list.

\begin{table}[t]
      \centering
      \begin{tabular}{>{\raggedright\arraybackslash\ttfamily\scriptsize}p{0.46\textwidth}<{}}
          \toprule

\headercolorlong
\textbf{top-$k$ Candidate Search Diagnostic Block}\\
Fine-grained Diagnostic Feedback:\\
The diagnostic model ranks the following line as suspicious:\\
Suspicious Line:\\
<BRANCH\_SPECIFIC\_SUSPICIOUS\_LINE>\\
This line may be relevant to the observed execution failure.\\

\bottomrule
      \end{tabular}
      \caption{Branch-specific diagnostic feedback block used in top-$k$ Candidate Search. Each branch uses the same diagnostic-guided refinement prompt but substitutes a different top-$k$ suspicious line.}
      \label{tab:topk_candidate_search_prompt}
\end{table}

Each branch generates one candidate revision. Passing candidates terminate the loop; otherwise, failed candidates are ranked by execution outcomes to choose the next program state.

\section{Detailed Failure Case Analysis}
\label{sec:appendix_failure_cases}

This section provides representative LiveCodeBench failures for the three categories in Section~\ref{sec:analysis}. 
These examples illustrate what remains difficult after diagnostic feedback: the language model must still infer the right algorithm, interpret the task correctly, and maintain long-range program state.

\paragraph{Language Model Reasoning Failure.}
Table~\ref{tab:failure_case_pattern1} shows a purchasing problem where buying $k$ units of product $i$ costs $k^2P_i$. 
A correct solution must reason about marginal costs and purchase units in increasing marginal-cost order. 
The generated solution instead binary searches the target number of units and uses a naive purchasing loop that does not correctly test whether that target is achievable. 
This failure reflects a missing algorithmic insight rather than a local implementation mistake.

\paragraph{Misinterpretation of Task Requirements.}
Table~\ref{tab:failure_case_pattern2} shows a problem about merging whole bags of stones and counting possible final XOR values. 
The generated solution builds a linear basis over the initial stone counts, which is appropriate for subset-XOR problems but not for this merge operation. 
The model recognizes a relevant mathematical keyword, XOR, but applies the wrong problem formulation.

\paragraph{Long-Range State and Simulation Errors.}
Table~\ref{tab:failure_case_pattern3} shows a simulation problem where adults give stones over time. 
A correct solution must track which adults still have stones after each year. 
The generated solution collapses this history into a single counter, \texttt{total\_stones}, so it cannot represent adults running out of stones. 
This illustrates failures where the code loses information needed across multiple state transitions.

\begin{table*}[]
\centering
\small
\renewcommand{\arraystretch}{1.3}
\begin{tabularx}{\textwidth}{@{} p{2cm} p{6.5cm} X @{}}
\toprule
\textbf{Failure Pattern} & \textbf{Original Task Description} & \textbf{Generated Failure Case Snippet} \\
\midrule
\textbf{Language Model Reasoning Failure} 
& \textbf{Task:} There are \(N\) types of products, each having \(10^{100}\) units in stock. You may buy any non-negative integer number of units of each product, and buying \(k\) units of the \(i\)-th product costs \(k^{2}P_i\) yen. Given a budget of at most \(M\) yen for the total purchase across all product types, determine the maximum possible total number of units you can buy (summing over all types).
& 
\vspace{-3mm}
{\tiny \ttfamily
\begin{tabular}{@{}l@{}}
def main(): \\
\quad import sys \\
\quad input = sys.stdin.read \\
\quad data = input().split() \\
\quad N = int(data[0]) \\
\quad M = int(data[1]) \\
\quad P = list(map(int, data[2:N+2])) \\
\quad def can\_buy(x): \\
\quad \quad total\_cost = 0 \\
\quad \quad for price in P: \\
\quad \quad \quad k = 0 \\
\quad \quad \quad while True: \\
\quad \quad \quad \quad cost = (k + 1) * (k + 1) * price \\
\quad \quad \quad \quad if total\_cost + cost $>$ M: \\
\quad \quad \quad \quad \quad break \\
\quad \quad \quad \quad total\_cost += cost \\
\quad \quad \quad \quad k += 1 \\
\quad \quad \quad \quad if total\_cost $>$ M: \\
\quad \quad \quad \quad \quad return False \\
\quad \quad \quad if total\_cost $>$ M: \\
\quad \quad \quad \quad return False \\
\quad \quad return total\_cost $<$= M \\
\quad low, high = 0, 10**9 \\
\quad answer = 0 \\
\quad while low $<$= high: \\
\quad \quad mid = (low + high) // 2 \\
\quad \quad if can\_buy(mid): \\
\quad \quad \quad answer = mid \\
\quad \quad \quad low = mid + 1 \\
\quad \quad else: \\
\quad \quad \quad high = mid - 1 \\
\quad print(answer) \\
if \_\_name\_\_ == "\_\_main\_\_": \\
\quad main()
\end{tabular}
} \\
\bottomrule
\end{tabularx}
\caption{Failure case (Pattern 1): Language Model Reasoning Failure.}
\label{tab:failure_case_pattern1}
\end{table*}

\begin{table*}[]
\centering
\small
\renewcommand{\arraystretch}{1.3}
\begin{tabularx}{\textwidth}{@{} p{2cm} p{6.5cm} X @{}}
\toprule
\textbf{Failure Pattern} & \textbf{Original Task Description} & \textbf{Generated Failure Case Snippet} \\
\midrule
\textbf{Misinterpretation of Requirements} 
& \textbf{Task:} There are \(N\) bags labeled \(1,2,\ldots,N\), where bag \(i\) initially contains \(A_i\) stones. You may repeat the following operation any number of times (possibly zero): choose two bags \(A\) and \(B\), and move all stones from bag \(A\) into bag \(B\) (so bag \(A\) becomes \(0\) and bag \(B\) increases accordingly). Let \(B_i\) denote the final number of stones in bag \(i\); consider the value \(B_1 \oplus B_2 \oplus \cdots \oplus B_N\), where \(\oplus\) is bitwise XOR (defined by binary digits: the \(2^k\) bit is \(1\) iff exactly one operand has a \(1\) at that bit, and XOR over multiple integers is taken associatively). Under the given setting it can be proved that only finitely many XOR values are achievable; compute how many distinct values can occur.
& 
\vspace{-3mm}
{\tiny \ttfamily
\begin{tabular}{@{}l@{}}
def main(): \\
\quad import sys \\
\quad input = sys.stdin.read \\
\quad data = input().split() \\
\quad N = int(data[0]) \\
\quad A = list(map(int, data[1:N+1])) \\
\quad basis = [] \\
\quad for number in A: \\
\quad \quad for b in basis: \\
\quad \quad \quad number = min(number, number \^{} b) \\
\quad \quad if number $>$ 0: \\
\quad \quad \quad basis.append(number) \\
\quad \quad \quad basis.sort(reverse=True) \\
\quad print(1 $<$$<$ len(basis)) \\
if \_\_name\_\_ == "\_\_main\_\_": \\
\quad main()
\end{tabular}
} \\
\bottomrule
\end{tabularx}
\caption{Failure case (Pattern 2): Misinterpretation of Requirements.}
\label{tab:failure_case_pattern2}
\end{table*}

\begin{table*}[]
\centering
\small
\renewcommand{\arraystretch}{1.3}
\begin{tabularx}{\textwidth}{@{} p{2cm} p{6.5cm} X @{}}
\toprule
\textbf{Failure Pattern} & \textbf{Original Task Description} & \textbf{Generated Failure Case Snippet} \\
\midrule
\textbf{Long-Range State and Simulation Errors} 
& \textbf{Task:} On a certain planet there are \(N\) aliens, all initially minors; the \(i\)-th alien currently has \(A_i\) stones and will become an adult exactly \(i\) years later. Whenever an alien becomes an adult, every alien who is already an adult and has at least one stone gives exactly one stone as a congratulatory gift to the alien who has just become an adult. Assuming no new aliens will be born, determine how many stones each alien has after \(N\) years.
& 
\vspace{-3mm}
{\tiny \ttfamily
\begin{tabular}{@{}l@{}}
def main(): \\
\quad import sys \\
\quad input = sys.stdin.read \\
\quad data = input().split() \\
\quad N = int(data[0]) \\
\quad A = list(map(int, data[1:N+1])) \\
\quad B = A[:] \\
\quad total\_stones = 0 \\
\quad for i in range(N): \\
\quad \quad B[i] += total\_stones \\
\quad \quad if B[i] $>$ 0: \\
\quad \quad \quad total\_stones += 1 \\
\quad \quad \quad B[i] -= 1 \\
\quad print(" ".join(map(str, B))) \\
if \_\_name\_\_ == "\_\_main\_\_": \\
\quad main()
\end{tabular}
} \\
\bottomrule
\end{tabularx}
\caption{Failure case (Pattern 3): Long-Range State and Simulation Errors.}
\label{tab:failure_case_pattern3}
\end{table*}

\begin{table*}[t]
\centering
\small
\begin{tabular}{p{0.18\textwidth} p{0.76\textwidth}}
\toprule
\textbf{Item} & \textbf{Content} \\
\midrule
\textbf{Task Description}
&
Given $N$ foods and a maximum total calorie limit $X$, each food provides a specific amount of exactly one type of vitamin, Vitamin 1, 2, or 3, and costs a certain number of calories. The goal is to choose a subset of foods such that the total consumed calories do not exceed $X$, while maximizing the minimum total intake among the three vitamins. In other words, the solution should balance the intake of all three vitamins and make the lowest accumulated vitamin amount as large as possible. \\
\midrule
\textbf{Initial Code}
&
\begin{minipage}[t]{0.76\textwidth}
\begin{lstlisting}[language=Python, basicstyle=\ttfamily\scriptsize, breaklines=true]
def main():
    import sys
    input = sys.stdin.read
    data = input().splitlines()
    N, X = map(int, data[0].split())
    foods = [tuple(map(int, line.split())) for line in data[1:N+1]]
    dp = [[0] * (X + 1) for _ in range(4)]
    for v, a, c in foods:
        for j in range(X, c - 1, -1):
            dp[v][j] = max(dp[v][j], dp[v][j - c] + a)
    max_min_intake = 0
    for calories in range(X + 1):
        min_intake = min(dp[1][calories], dp[2][calories], dp[3][calories])
        max_min_intake = max(max_min_intake, min_intake)
    print(max_min_intake)
if __name__ == "__main__":
    main()
\end{lstlisting}
\end{minipage}
\\
\midrule
\textbf{Predicted Suspicious Line}
&
\begin{minipage}[t]{0.76\textwidth}
\begin{lstlisting}[language=Python, basicstyle=\ttfamily\scriptsize, breaklines=true]
dp = [[0] * (X + 1) for _ in range(4)]
\end{lstlisting}
\end{minipage}
\\
\midrule
\textbf{Final Passing Candidate}
&
\begin{minipage}[t]{0.76\textwidth}
\begin{lstlisting}[language=Python, basicstyle=\ttfamily\scriptsize, breaklines=true]
def main():
    import sys
    input = sys.stdin.read
    data = input().splitlines()
    N, X = map(int, data[0].split())
    foods = [tuple(map(int, line.split())) for line in data[1:N+1]]
    vitamin_foods = {1: [], 2: [], 3: []}
    for v, a, c in foods:
        vitamin_foods[v].append((a, c))
    def can_achieve(m):
        dp = [False] * (X + 1)
        dp[0] = True
        for v in range(1, 4):
            new_dp = [False] * (X + 1)
            for a, c in vitamin_foods[v]:
                if a >= m:
                    for j in range(X, c - 1, -1):
                        if dp[j - c]:
                            new_dp[j] = True
            dp = new_dp
        return any(dp)
    low, high = 0, max(a for v, a, c in foods)
    answer = 0
    while low <= high:
        mid = (low + high) // 2
        if can_achieve(mid):
            answer = mid
            low = mid + 1
        else:
            high = mid - 1
    print(answer)
if __name__ == "__main__":
    main()
\end{lstlisting}
\end{minipage}
\\
\bottomrule
\end{tabular}
\caption{Example of {\method} from LiveCodeBench. {\method} correctly solves the task with refinement that uses a binary search over a target value $m$ and checks whether one subset of foods can make all three vitamin totals at least $m$ within budget $X$.}
\label{tab:case-study-livecodebench}
\end{table*}

\begin{table*}[t]
\centering
\small
\begin{tabular}{p{0.10\textwidth} p{0.62\textwidth} p{0.22\textwidth}}
\hline
\textbf{Method} & \textbf{Failed Code Snippet} & \textbf{Failure Analysis} \\
\hline
Self-Debug &
\begin{minipage}[t]{\linewidth}
\begin{lstlisting}[language=Python,basicstyle=\ttfamily\scriptsize,breaklines=true]
def main():
    import sys
    from collections import defaultdict
    input = sys.stdin.read
    data = input().splitlines()
    N, X = map(int, data[0].split())
    foods = [tuple(map(int, line.split())) for line in data[1:N+1]]
    vitamin_foods = defaultdict(list)
    for v, a, c in foods:
        vitamin_foods[v].append((a, c))
    dp = [[0] * (X + 1) for _ in range(4)]
    for v in range(1, 4):
        for a, c in vitamin_foods[v]:
            for cal in range(X, c - 1, -1):
                dp[v][cal] = max(dp[v][cal], dp[v][cal - c] + a)
    low, high = 0, min(dp[1][X], dp[2][X], dp[3][X]) + 1
    while low < high:
        mid = (low + high) // 2
        can_achieve = False
        for cal in range(X + 1):
            if dp[1][cal] >= mid and dp[2][cal] >= mid and dp[3][cal] >= mid:
                can_achieve = True
                break
        if can_achieve:
            low = mid + 1
        else:
            high = mid
    print(low - 1)
if __name__ == "__main__":
    main()
\end{lstlisting}
\end{minipage}
&
The solution computes three independent knapsack tables, one for each vitamin type. However, the feasibility check is incorrect: it checks whether there exists the same calorie value \texttt{cal} such that all three independent DP tables reach the target value. This does not model the shared calorie budget. The correct condition should check whether there exist calorie allocations $c_1, c_2, c_3$ such that $c_1+c_2+c_3 \leq X$ and each vitamin reaches the target. \\
\hline
NL-Debugging &
\begin{minipage}[t]{\linewidth}
\begin{lstlisting}[language=Python,basicstyle=\ttfamily\scriptsize,breaklines=true]
def main():
    import sys
    input = sys.stdin.read
    data = input().splitlines()
    N, X = map(int, data[0].split())
    foods = []
    for i in range(1, N + 1):
        V, A, C = map(int, data[i].split())
        foods.append((V, A, C))
    def can_achieve(min_vitamin):
        dp = [[0] * 4 for _ in range(X + 1)]
        for V, A, C in foods:
            for c in range(X, C - 1, -1):
                dp[c][V] = max(dp[c][V], dp[c - C][V] + A)
        return dp[X][1] >= min_vitamin and dp[X][2] >= min_vitamin and dp[X][3] >= min_vitamin
    low, high = 0, 200000
    answer = 0
    while low <= high:
        mid = (low + high) // 2
        if can_achieve(mid):
            answer = mid
            low = mid + 1
        else:
            high = mid - 1
    print(answer)
if __name__ == "__main__":
    main()
\end{lstlisting}
\end{minipage}
&
The DP state is invalid for the problem. \texttt{dp[c][v]} only records the best amount of one vitamin under calorie budget \texttt{c}; it does not represent a joint selection of foods. Therefore, \texttt{dp[X][1]}, \texttt{dp[X][2]}, and \texttt{dp[X][3]} may correspond to three different subsets of foods, each using up to $X$ calories independently. The algorithm may thus falsely conclude that all vitamin requirements are simultaneously feasible under the single shared budget $X$. \\
\hline
Self-Refine &
\begin{minipage}[t]{\linewidth}
\begin{lstlisting}[language=Python,basicstyle=\ttfamily\scriptsize,breaklines=true]
def main():
    import sys
    input = sys.stdin.read
    data = input().splitlines()
    N, X = map(int, data[0].split())
    foods = []
    for i in range(1, N + 1):
        V, A, C = map(int, data[i].split())
        foods.append((V, A, C))
    dp = [[0] * (X + 1) for _ in range(4)]
    for V, A, C in foods:
        for c in range(X, C - 1, -1):
            dp[V][c] = max(dp[V][c], dp[V][c - C] + A)
    max_min_vitamin = 0
    for c in range(X + 1):
        min_vitamin = min(dp[1][c], dp[2][c], dp[3][c])
        max_min_vitamin = max(max_min_vitamin, min_vitamin)
    print(max_min_vitamin)
if __name__ == "__main__":
    main()
\end{lstlisting}
\end{minipage}
&
The solution keeps the same flawed per-vitamin DP formulation after refinement. It computes \texttt{dp[V][c]} independently for each vitamin and then takes \texttt{min(dp[1][c], dp[2][c], dp[3][c])} at the same calorie value \texttt{c}. These values may come from different food subsets and do not guarantee a jointly feasible selection under the shared budget \(X\). Thus, the code can overestimate the achievable minimum vitamin intake. \\
\hline
\end{tabular}
\caption{Case study of failed solutions produced by Self-Debug, NL-Debugging, and Self-Refine. All failures arise from an incorrect modeling of the shared calorie budget across three vitamin categories.}
\label{tab:case-study-failed-debugging}
\end{table*}


\end{document}